\documentstyle[12pt,epsfig]{article}
\begin{document}
\def\affil#1{\vspace*{-2.5ex}{\topsep 0pt\center\small\it#1\endcenter}}
\def\slash#1{\setbox0=\hbox{$#1$}#1\hskip-\wd0\hbox to\wd0{\hss\sl/\/\hss}}
%beginning of source
%  from Kopka, ch. 7.3.3
\newcounter{saveeqn}
\newcommand{\alpheqn}{\setcounter{saveeqn}{\value{equation}}%
\stepcounter{saveeqn}\setcounter{equation}{0}%
\renewcommand{\theequation}{\mbox{\arabic{saveeqn}\alph{equation}}}}
\newcommand{\reseteqn}{\setcounter{equation}{\value{saveeqn}}
\renewcommand{\theequation}{\arabic{equation}}}
%end of source
\baselineskip=24pt

\title{Study of Quark Propagator Solutions to the Dyson--Schwinger Equation in
a Confining Model}
\author{Douglas W. McKay and Herman J. Munczek  \\ {\small\it Department of
Physics and Astronomy, The University of Kansas, Lawrence, KS 66045}}

\maketitle

\begin{abstract}
\baselineskip=24pt
We solve the Dyson--Schwinger equation for the quark propagator in a model with
singular infrared behavior for the gluon propagator. We require that the
solutions, easily found in configuration space, be tempered distributions and
thus have Fourier transforms. This severely limits the boundary conditions
that the solutions may satisify. The sign of the dimensionful parameter that
characterizes the model gluon propagator can be either positive or negative. If
the sign is negative, we find a unique solution. It is singular at the origin
in momentum space, falls off like $1/p^2$ as $p^2\rightarrow +/-\infty$, and it
is truly nonperturbative in that it is singular in the limit that the
gluon--quark interaction approaches zero. If the sign of the gluon propagator
coefficient is positive, we find solutions that are, in a sense that we
exhibit, unconstrained linear combinations of advanced and retarded
propagators. These solutions are singular at the origin in momentum space,
fall off like $1/p^2$ asympotically, exhibit ``resonant--like" behavior at the
position of the bare mass of the quark when the mass is large compared to the
dimensionful interaction parameter in the gluon propagator model, and smoothly approach a
linear combination of, free--quark, advanced and retarded two--point functions
in the limit that the interaction approaches zero. In this sense, these solutions behave in an increasingly
``particle--like" manner as the quark becomes heavy. The Feynman propagator and
the Wightman function are not tempered distributions and therefore are not
acceptable solutions to the Schwinger--Dyson equation in our model. On this
basis we advance
several arguments to show that the Fourier--transformable solutions we find
are consistent with quark confinement, even though they have singularities on
the real $p^2$--axis.
\end{abstract}

\section{Introduction}

A classic approach to understanding the behavior of confined particles is to
model and solve the Dyson-Schwinger (DS) equations for the particles'
propagators.  Since confinement is generally regarded as an infrared
phenomenon, the emphasis is naturally on the infrared region of the kernels of
the DS equations. Taking clues from studies of the infrared behavior of
propagators in pure Yang--Mills theory, one can adopt a vector-meson propagator
model motivated by such studies and insert it in the kernel for the fermionic
propagator equation and study issues such as fermion confinement, chiral
symmetry breaking, the interplay between the scales for these two
phenomena, and gauge dependence of solutions. There are two
extreme views of the infrared behavior of the gluon propagator. One is that the
singularity at $q^2=0$ is much stronger than the $1/q^2$ behavior of the
perturbative propagator, with variants of $1/q^4$ often proposed, and
the other, in complete contrast, is that the propagator vanishes as
$q^2\rightarrow0$.
Because of the wide and rather successful application of the former type of
behavior to bound state problems, we will adopt a frequently studied model of
this type, proposed some time ago by one of the us\cite{{hman},{hm}}, for our
analysis. 

Intuition for the interpretation and application of quantum field theories is
built upon an intimate interplay between configuration space and momentum space
considerations. The interaction Lagrangian and its symmetry properties are studied
in configuration space, and space--time boundary conditions of the Green
functions of the theory are crucial to their interpretation. On the other
hand, the particle spectrum and the scattering and decay processes contained in the
theory are more intuitively assessed in momentum space. The particle content is
revealed in the Green functions  by their branch cuts and poles in momentum
space. %\cite{jc} 
The complementarity of
the configuration space and momentum space views is especially clear in the
interchangeability of the terms short--distance and long--distance with
ultraviolet and infrared (or  hard and soft) to describe the physics of a
situation. It is not surprising, therefore, that we assume that Green
functions in (Minkowsky) configuration space have Fourier
transforms to momentum space and vice--versa. Indeed, this property is at the
foundation of the standard approach to particles and fields.

We return to the study of fermion propagators in the infrared domain to see
what insight can be gained by requiring that the solutions to the DS equation
be Fourier transformable. The question is not idle, since several 
solutions proposed in the literature as possible models of confined behavior do
not have Fourier transforms\cite{{hm},{cb},{dc}}. This consideration is one of the 
motivations for the present work. Important collateral questions that
will occupy us are those of the propagator behavior in the large mass limit,
the asymptotic behavior in both timelike and spacelike directions in momentum
space, and the behavior of the propagator in the limit that the infrared
``gluon--fermion" interaction is turned off. We choose a simple enough model
that ``abelianized" Ward--Takahashi identities can be enforced at the
fermion--gauge boson vertex and still leave us with a model whose propagator we
can solve for exactly and 
whose
Fourier transform we can evaluate. Perhaps unique to the present study is that we
remain strictly in Minkowski space in setting up and solving our model
equations.\footnote{With care taken to handle the continuation to Minkowski
space 
properly, an equivalent Euclidean space treatment can be given.}

\section{Defining the Model and Solving for the Propagator}

We begin by developing the model for our study of Fourier transformable
solutions to the fermion Dyson--Schwinger equation. 
Our crucial ingredient for the DS equation is the infrared gluon propagator
model\cite{{hman},{hm}} used a number of times since\cite{{cb},{dc},{pjhm}}. As mentioned
in the 
introduction, a number of studies of the gluon propagator suggest\cite{mb}
that 
\begin{equation}                                                          
D(q^2) \rightarrow  \frac{\mu^2}{q^4}\qquad {\rm as}\; q^2\rightarrow 0
\end{equation}
where, in Landau Gauge, 
\begin{equation}
D_{\mu\nu}(q^2)=\Big(-g_{\mu\nu}+\frac{q_\mu q_\nu}{q^2}\Big) D(q^2).
\end{equation}
The Fourier transform of $1/(q^2+i\epsilon)^2$ does not exist\cite{img},
however, and a regularization must be prescribed to define a gluon 
propagator that has a Fourier transform. Defining $D^{(\lambda)}(q^2)$ as
\begin{equation}
D^{(\lambda)}(q^2)=\frac{1}{\Gamma(-\lambda)}\frac{M}{(q^2+i\epsilon)^\lambda },
\end{equation}
where $M$ has dimensions of (mass)$^{2(\lambda-1)}$, the limit\cite{hman}
\begin{equation}
\lim\limits_{\lambda\rightarrow 2} D^{(\lambda)}_{\mu\nu}(q^2)
=i\mu^2\delta^4(q)(2\pi)^4g_{\mu\nu},
\end{equation}

\noindent where $\mu^2$ can be positive or negative, defines the propagator for our infrared DS equation
study.\footnote{S. 
Blaha\cite{sb} studied $PP(1/(q^2+i\epsilon)^2)$, but in a perturbative context. 
Pagels studied an alternative prescription, with a quark propagator 
vanishing as $\lambda\rightarrow 2$, to handle the $1/q^4$
singularity. This leads to a different DS Equation from ours[9]. }
The simple form of the confining propagator then is
\begin{equation}
D^{\mu\nu}(x)=i\mu^2g^{\mu\nu} 	
\end{equation}
in configuration space.\footnote{The configuration space propagator (5) is
consistent with confinement, since it 
clearly does not satisfy the cluster decomposition property[10].}

The general form of the DS equation is, in momentum space, 
\begin{equation}
1=(\slash p -m) S(p)-i \int\frac{d^4k}{(2\pi)^4}\gamma_\mu
D^{\mu\nu}(k)\Lambda_\nu (p+k,p),
\end{equation}
where $\Lambda_\nu$ is the dressed vertex defined in Appendix B. Inserting our model propagator (4),
we obtain 
\begin{equation}
1=(\slash p - m)S(p)+\mu^2\gamma^\mu\cdot \Lambda_\mu(p,p).
\end{equation}
We assume that $\Lambda_\mu(p+k,p)$ obeys the Ward identity
\begin{equation}
\Lambda_\mu(p,p)=-{\partial\over\partial p^\mu}S(p),
\end{equation}
which is exact in an abelian gauge theory and true also in non-abelian gauge
theory if the ghost contributions to the Ward--Takahashi identity are 
of order $k$ and higher. The DS equation (7) then reads
\begin{equation}
1=(\slash p-m)S(p)-\mu^2\gamma^\mu\cdot {\partial\over\partial p^\mu}S(p),
\end{equation} in our model. Taking the Fourier transform of Eq. (9), we find
that the $x$--space DS equation
that forms the basis for the present investigation is
\begin{equation}
\delta^4(x)=(i\slash\partial -m)S(x)+ i\mu^2\gamma\cdot x S(x).
\end{equation} 
Phenomenologically, the dimensionful parameter $\mu$ can be expressed in
terms of the QCD 
hadron dynamics scale by fitting the pion decay constant, for example. 

Equation (10) has the factorizable solution\cite{hm}
\begin{equation}                        
S(x)=e^{-\mu^2x^2/2} S_0(x),
\end{equation}
where $S_0(x)$ is a solution of the free DS equation. 
In order to avoid an exponential blow--up that invalidates the Fourier
transform, we must choose an appropriate $S_0(x)$.\footnote{Momentum space
solutions to this model that have been offered in the literature [2,3,4] are not Fourier
transformable.} For the case $\mu^2>0$,
this 
means that the free propagator choice must be\cite{srp}
\begin{equation}
\bar S_0(x) = -\Big(i\slash\partial
+m\Big)\Big(1+C\epsilon(x_0)\Big)\bar\Delta(x^2), 
\end{equation}
where $C$ is an arbitrary constant and $\epsilon(x_0)\bar\Delta(x^2)$ obeys
the free homogeneous DS equation. The choices $C=\mp 1$ yield advanced and
retarded Green functions, respectively. $\bar\Delta(x^2)$ obeys the
inhomogeneous Klein--Gordon equation and has the form[11]
\begin{equation}
\bar\Delta(x^2)=\frac{1}{4\pi}\Big(\delta(x^2)-\frac{m^2}{2}\theta(x^2)
\frac{J_1\Big(m\sqrt{x^2}\big)}{m\sqrt{x^2}}\Big),
\end{equation} and $\theta(x^2)$ prohibits the $x^2<0$ region where, otherwise, 
the full
propagator Eq. 
(11) would blow up. 
For the $\mu^2<0$ case in Eq. (4),
one needs a solution with $\theta(-x^2)$, which follows from Eq. (13) by
adding the appropriate solution to the homogeneous equation;
namely\footnote{This solution is the unique one in which $\theta(-x^2)$
appears, 
as required in the $\mu^2<0$ case. That is, for this case, $C=0$.}

\begin{eqnarray}
\tilde\Delta(x^2) & = &
\bar\Delta(x^2)+\frac{m^2}{4\pi}\frac{J_1(m\sqrt{x^2})}{m\sqrt{x^2}} \\
& = &
\frac{1}{4\pi}\Big(\delta(x^2)+\frac{m^2}{2}\theta(-x^2)\frac{J_1\Big(m\sqrt{x^2}\big)}
{m\sqrt{x^2}}\Big),\nonumber \\
\nonumber
\end{eqnarray}
with $\tilde S_0(x)=-(i\slash\partial +m)\tilde\Delta(x^2)$. Note that Eq. (13)
represents a tempered distribution, and therefore has a Fourier transform, while
Eq. (14) does not.
We emphasize that a Feynman propagator is {\it not} an acceptable 
choice for $S_0(x)$ in Eq. (11), because the corresponding solution, $S(x)$,
does not have a Fourier transform.

In summary, we have the two cases
\alpheqn\begin{equation}
\bar S(x)=e^{-\frac{\mid\mu^2\mid}{2}x^2}\bar S_0(x)
\end{equation}
and
\begin{equation}
\tilde S(x)=e^{\frac{\mid\mu^2\mid}{2}x^2}\tilde S_0(x),
\end{equation}
\reseteqn

\noindent corresponding to the choices $\mu^2>0$ and $\mu^2<0$, respectively,
for the gluon propagator model 
 in Eq. (4). The non--interacting Green--functions $\bar
S_0(x)$ and $\tilde S_0(x)$\footnote{$\tilde S_0(x)$ does not have a Fourier
transform. Thus
(15b) is a truly nonperturbative solution. As we will see below, though
$\tilde S_0(x^2)$ is the $\mu\rightarrow 0$ limit of (15b), the Fourier
transform of (15b)
is singular as $\mu\rightarrow 0$.} 
are manifestly regained in the $\mu\rightarrow 0$
limit. Avoiding the exponential blow-up as $x^2\rightarrow -\infty$ in Eq.
(15a) and as $x^2\rightarrow + \infty$ in Eq. (15b) dictates the choices of
$\bar\Delta (x)$ and $\tilde\Delta(x)$ in Eqs. (13) and
(14), as necessary conditions to ensure that Fourier transforms to 
momentum space exist.

In our model, the Wightman function, $S_W(x)\equiv <0\mid\psi(x)\bar\psi(0)\mid
0>$, could be identified plausibly as
\begin{equation}
S_W(x)=e^{-{\mu^2\over 2}x^2}(i\slash\partial +m)W_0(x),
\end{equation}                                      
where $W^0(x)$ is the free--field scalar
Wightman function\cite{srp}
\begin{equation}
W^0(x)={m\over 4\pi\sqrt{x^2}}K_1(\sqrt{x^2}m),
\end{equation}
in terms of a standard Hankel function. If Wightman functions are tempered
distributions one can prove  that free 
fermion asymptotic states exist\cite{prev}. $S_W(x)$ as defined above is not a
tempered distribution, so our model is consistent with fermion 
confinement. The Schwinger model, which is solvable, illustrates such a
connection between pathologies of the Wightman functions and confinement. The
fermion Wightman functions in Coulomb gauge blow up exponentially in
configuration space and fermion
states, correspondingly, do not appear in the spectrum\cite{jljs}.

Let us now take up the evaluation of the Fourier transforms of Eqs. (15a) and
(15b) and examine their behavior in momentum space. The essential calculations
that must be performed are the Fourier transforms of
$e^{-\mid\mu^2\mid x^2}\bar\Delta(x^2)$ 
and $e^{+\mid\mu^2\mid x^2}\tilde\Delta(x^2)$; namely, choosing
$e^{-\mid\mu^2\mid x^2}\bar\Delta(x^2)$ for discussion, we have, adopting the
convention $\mu^2>0$

\begin{equation}
\bar B(p^2)=-m\int d^4xe^{ip\cdot x}e^{-{\mu^2\over 2} x^2}\bar\Delta
(x^2)\Big(1+C\epsilon(x_0)\Big) ,
\end{equation}
and 
\begin{eqnarray}
p^2\bar A(p^2) & = & -i\int d^4xe^{ip\cdot 
x}e^{-{\mu^2\over 2}x^2}\slash p\slash\partial\Big[\bar\Delta(x^2)\Big(1+C\epsilon(x_0) 
\Big)\Big]\nonumber \\
& = & \Big(p^2+\mu^2p\cdot\partial_p\Big)\bar B(p^2)/m. \\
\nonumber
\end{eqnarray} We have defined $\bar A(p^2)$ and $\bar B(p)$ in terms of
$\bar S(p)$ to be 
\begin{equation}
\bar S(p)=\slash p \bar A(p^2)+\bar B(p^2)=\int d^4xe^{ip\cdot x}\bar S(x).
\end{equation} Details of our evaluation of the Fourier transform (18) are
given in the Appendix A, where we present a procedure that can be applied to
any function $F(x^2)$ which has a one dimensional (in $x^2$) Fourier transform.
We also show there how to choose contours that give
improved convergence for the numerical evaluation of the Fourier transform for
timelike and spacelike values of the momentum--space argument. Writing the
$C=0$ result for $\bar B(p^2)$ derived in Appendix A in the form 

\begin{equation}
\frac{\bar B(p^2)}{m}=\frac{i}{2}\int\limits^\infty_{-\infty} d\nu
\epsilon(\nu)e^{-ip^2\nu+i{m^2\nu\over 1-2i\mu^2\nu}},
\end{equation} we factor out $e^{-{m^2\over 2\mu^2}}$ and introduce a 
variable, $\lambda^2$, as follows:

\begin{eqnarray}
e^{i{m^2\nu\over 1-2 i\mu^2\nu}}& = & e^{-{m^2\over
2\mu^2}}\int\limits^\infty_{-\infty}d\tau\delta(\nu-\tau)e^{{m^2\over
2\mu^2}{1\over 1-2i \mu^2\tau}}\nonumber \\
& = & e^{-{m^2\over 2\mu^2}}\int\limits^{\infty}_{-\infty}{d\tau\over 2\pi}
\int\limits^{\infty}_{-\infty} d\lambda^2e^{i\lambda^2(\nu-\tau)}e^{{m^2\over
2\mu^2}{1\over 1-2 i\mu^2\tau}} \\
\nonumber
\end{eqnarray}             
Substituting (22) into (21), exchanging the order of $\lambda^2$ and $\tau$
integration and evaluating the integral over $\nu$ produces
\begin{equation}
{\bar B\over m}(p^2)=PP\int\limits^\infty_{-\infty}d\lambda^2
{\sigma(\lambda^2)\over 
p^2-\lambda^2}\; ,
\end{equation}
where

\begin{eqnarray}
\sigma(\lambda^2)& = & e^{-{m^2\over 2\mu^2}}\int\limits^\infty_{-\infty}
{d\tau\over 2\pi}e^{-i\lambda^2\tau+{m^2\over 2\mu^2}{1\over1-2 i\mu^2\tau}}
\nonumber \\
& = & e^{-m^2\over2\mu^2}\Bigg(\delta(\lambda^2)+\theta(\lambda^2)
{m\over2\mu^2\lambda}e^{-{\lambda^2\over 2\mu^2}}I_1\Big(
{\lambda m\over \mu^2}\Big)\Bigg)\\
& \equiv & e^{-m^2\over2\mu^2}\delta(\lambda^2)+\bar\sigma(\lambda^2)
\theta(\lambda^2)\; .\nonumber \\
\nonumber
\end{eqnarray}
Thus our representation for $C=0$ is

\alpheqn\begin{equation}
\frac{\bar B(p^2)}{m}=PP\Big[{e^{-{m^2\over 2\mu^2}}\over p^2}+\int\limits^\infty_0
d\lambda^2{\bar\sigma(\lambda^2)\over p^2-\lambda^2}\Big]\; .
\end{equation}                                     
Equation (23) represents $\bar B(p^2)$ as a superposition of free propagators of
mass $\lambda$. The Fourier transform of $\epsilon(x_0)\bar\Delta(x^2,\lambda)$
is obtained from that of $\bar\Delta(x^2,\lambda)$ by the substitution
$(p^2-\lambda^2)^{-1}\rightarrow -i\pi\epsilon(p_0)\delta(p^2-\lambda^2)$. 
Therefore, the Fourier transform of the second term in Eq. (18) is
\begin{equation}
C\Big[ i\pi\epsilon (p_0)e^{-{m^2\over
2\mu^2}}\Bigg(\delta(p^2)+\theta(p^2){m\over
2\mu^2}{1\over\sqrt{p^2}}I_1\Big({m\sqrt{p^2}\over
\mu^2}\Big)\Bigg)e^{-p^2/2\mu^2}\Big].
\end{equation}
\reseteqn
Equations (25a,b) show 
several key features of the momentum space behavior of the propagator that is
the solution to the DS equation in our model, and we turn to discussion of
these points in the next section.

\section{Properties of the Fermion Propagator}

\subsection{The $\mu^2>0$ case }

The most obvious features of Eqs. (25a,b) are the singularities at $p^2=0$. As
we emphasize  in Sec.~4, these are not the singularitites of a Feynman
propagator. 
Next we note that if $\mu^2\not= 0$, the $\lambda^2$ integral clearly converges
since

\begin{eqnarray}
I_1(x)& \longrightarrow & {1\over \sqrt{2\pi x}}e^x\qquad  {\rm as}\; \; 
x\rightarrow \infty\nonumber \\
{\rm and}\; \; {1\over\lambda}e^{-{\lambda^2\over 2\mu^2}}I\Big({\lambda m\over
2\mu^2}\Big)& \rightarrow & \sqrt{{\mu\over\pi m}}{1\over\lambda^{3/2}}
e^{-{\lambda^2\over 2\mu^2}+{\lambda m\over\mu^2}}\nonumber \\
\nonumber
\end{eqnarray}
which is strongly convergent. Therefore the asympototic behavior of
$\bar B(p^2)$ is
\begin{equation}
\bar B (p^2)\longrightarrow {m\over p^2} \hskip .75in {\rm as}\; \;
p^2\rightarrow \pm \infty\; ,\nonumber 
\end{equation} and the free--propagator ultraviolet behavior is
reproduced.\footnote{In fact, expanding Eq. (21) in powers of ${1\over p^2}$
shows that the asympototic behavior is given by $\bar B(p^2)/m\rightarrow
{1\over p^2-m^2}  +{\cal{O}}\Big({m^4\over p^6}\Big)$, as $p^2\rightarrow
\infty$.}

Next we consider the ${\mu^2\over m^2}\rightarrow 0$ limit of the expressions
(25a,b) for $\bar B(p^2)$.  The singularities at the origin vanish exponentially in this
limit. What happens to the principal part integral? The asymptotic expansion
for $I_1(x)$ shown above allows one to write the limit in the form

\begin{eqnarray}
\bar\sigma(\lambda^2)& \rightarrow &
{m\over 2\mu^2\lambda} {1\over \sqrt{2\pi}}{\mu\over
\sqrt{m\lambda}}e^{-{(\lambda-m)^2\over 2\mu^2}}\nonumber \\
& = &{\sqrt{m}\over \lambda^{3/2}}\delta(\lambda-m)\nonumber \\
\nonumber
\end{eqnarray}
which yields
\begin{equation}
\frac{\bar B(p^2)}{m}\rightarrow PP\Big({1\over
p^2-m^2}\Big)+i\pi C\epsilon(p_0)\delta(p^2-m^2),\qquad {\rm as}\;
\; 
{\mu^2\over m^2}\rightarrow 0\; ,\end{equation} which is the Fourier
transform of the free 
Green function $(1+C\epsilon(x_0))\bar\Delta(x^2)$ in Eqs. (12) and (13). This result establishes
that there is a smooth limit where the free momentum space Green function is
the Fourier transform of the free configuration space Green function. This
smooth limit does not obtain in our $\mu^2<0$ 
case of Eq. (4) (see Sec. (3.2) below), nor in the solutions
reported in the literature[2,3,4]. $\bar B(p^2)m$ is graphed for several
values of $\mu^2/ 
m^2$ in Fig. 1. The sharpening of the resonance--like behavior  at $p^2\cong m^2$
and the disappearance of the pole at $p^2= 0$ as 
$\mu^2/m^2\rightarrow 0$ is clearly
shown. Thus in the limit as the mass of the fermion becomes large compared to the scale
associated with the infrared behavior of the gluon propagator, the fermion
propagator becomes more and more particle--like, in the sense that it behaves
like $1/(p^2-m^2)$ everywhere. The pole at $p^2=0$ is {\it not} an actual  particle pole with the $i\epsilon$
prescription corresponding to a time ordered product 
 that insures unitarity in the perturbative expansion. This is true for any
value of $C$ in Eq. (12).

A blow--up of the region near $p^2=0$ for the $2\mu^2/m^2=0.2$ case is shown in
Fig. 2 to indicate just how sharp the pole is in this case where its weighting
factor is $e^{-m^2\over 2\mu^2}=e^{-5}$.

Figure 3 shows the value of $(\bar B(p^2)/m)(p^2-m^2)$ as a function of
$p^2/m^2$ for the
case $2\mu^2/m^2=0.2$. The rapid approach to the free Green function
behavior for large $p^2/m^2$is readily apparent.

\subsection{The Case $\mu^2<0$ -- an Example of a Singular
$\mu^2\rightarrow 0$ Limit}

The free Green function $\tilde \Delta(x^2),$ Eq. (14), and the corresponding
$\tilde S_0(x)$ are not tempered distributions and do not have Fourier
transforms. Nonetheless, the solution (15b) to the DS equation with the vertex
(8) and gluon infrared propagator (4) {\it does} have a Fourier transform
because the exponential factor $e^{\mid\mu^2\mid x^2}$ controls the
$\exp\Big(m\sqrt{\mid x^2 \mid}\Big)$ divergence of $J_1(m\sqrt{x^2})$ as
$x^2\rightarrow -\infty$. 

Following the same steps as before, one arrives at Eq. (21) but with the
opposite sign in front of
$\mu^2$. The representation corresponding to Eq. 25a is, 

\begin{equation}
\frac{\tilde B}{m} (p^2)=PP\Big[e^{{m^2\over2\mu^2}}{1\over p^2}+e^{{m^2\over
2\mu^2}}{m\over 2\mu^2}\int\limits^\infty_0 {d\lambda^2\over\lambda}{1\over
p^2+\lambda^2}e^{-{\lambda^2\over2\mu^2}}J_1\Big({m\lambda\over\mu^2}\Big)
\Big]\; .
\end{equation}

\noindent In the limit $p^2\rightarrow \pm \infty$, $\tilde B(p^2)/m\rightarrow
{1\over p^2}$ as in the previous case. In the limit $\mu^2/m^2\rightarrow0$,
there is no $\delta(\lambda-m)$ behavior, and there is no pole
at $p^2=m^2$. The whole expression {\it diverges} 
as $e^{{m^2\over 2\mu^2}}$ in the (singular) $\mu^2/m^2\rightarrow
0$ limit. The original free Green function $\tilde\Delta(x^2)$ is not Fourier
transformable, so the singular nature of the $\mu^2/m^2\rightarrow 0$ limit
merely reflects that fact.

\section{Discussion of Results and Conclusions}

We have reexamined a model for the infrared gluon propagator and quark-gluon
vertex previously discussed in the literature[1-5]. We found those solutions
to the 
quark propagator DS equation that admit Fourier transforms; we work directly in
Minkowski space. Such solutions lend themselves to the study of timelike and
spacelike behavior of the propagator without appeal to transformation to
Euclidean space and continuation of the solutions found to the timelike
region. The first solution ($\mu^2> 0$ case) presented has a smooth limit to a combination of
advanced 
and retarded Green functions of the free Dirac equation when the interaction
is turned off. In momentum space, the
real part of the free Green function is simply $PP(1/(p^2-m^2))$, which is also
the behavior of 
the full solution to the interacting model in the $\pm p^2\rightarrow 
\infty$ limit, as indicated in Fig. 3. The full propagator shows an interesting
particle--like behavior as the mass parameter grows large compared to the
infrared scale that characterizes the gluon propagator. This behavior is shown
in Fig. 1, and it gives an explicit picture of the increasingly
``free--particle--like" behavior expected as quarks become heavy. This is
particularly true of the top quark, of course. The solution has a 
 pole at $p^2=0$ that dominates when the quark mass parameter is of the
order of or less than the infrared scale in the gluon propagator, as shown in
Fig. 1. The singularity at the origin is suppressed by the
$\exp(-m^2/2\mu^2)$ factor displayed in Eqs. (25a,b) in the large quark mass
limit. 
Though the model is not completely realistic, since we do not include the
ultraviolet 
contribution from the gluon propagator, it does have the interesting feature
that, while the region near $p^2=0$ dominates when the quark mass is small, the
region near $p^2=m^2$ dominates as the quark mass gets
large. Regarding confinement, we note that the gluon propagator model, which is
constant in configuration space, clearly violates the cluster decomposition
property, which has been considered to be a sufficient condition for
confinement. For the quark propagator, absence of a singularity on the real 
$p^2$--axis is often taken to be a sufficient condition for 
confinement.                                      
In contrast, the quark propagator solution that we find is singular  at
the origin in momentum space. This singular behavior appears to be consistent
with confinement,  since the $i\epsilon$  prescription necessary to define Feynman diagrams
that build in the connection with unitary S--matrix elements between free
outgoing and incoming colored quark states is not permitted by the DS Eqs. (9)
or (10). Further reinforcing our point
about confinement is the fact that the most plausible candidate for the
two--point Wightman function in our model, Eq. (16), is not a tempered
distribution, a condition assumed for proving the existence of free--fermion
asympotic states.                              

The second Fourier transformable solution ($\mu^2<0$ case) that we presented
has the same ``free 
quark" asymptotic behavior in momentum space as the first, but the
particle--like resonant--behavior at $p^2=m^2$ in the large mass limit is
missing. Furthermore,  the large mass (or small interaction strength) limit is singular,
in keeping with the fact that, while this solution to the full, interacting DS
equation has a Fourier transform, the solution to the free equation does not.
When taken in momentum space, the limit to the free Green function simply does
not exist. The latter feature is shared by the solutions presented in [2],
[3] and [4], where the 
(Euclidean) momentum space propagators, extended to timelike values
of the argument, are not tempered distributions and
have no Fourier transform.

In conclusion, we have presented the tempered--distribution solutions to the DS equation for a simple
model in 4--dimensional Minkowsky space with a confining gluon propagator and a 
non--trivial quark gluon vertex. 
We offer the results as an interesting, instructive and useful addition to the
literature on the dynamics of confined quarks. 
\vskip .25in
\centerline{\Large Acknowledgements}
\medskip
\noindent We thank Pankaj Jain for discussions and assistance with the
graphs. The computational facilities of the Kansas Institute for Theoretical
and Computational Science were used for portions of this work. This research has
been supported in part by Department of Energy grant \#DE--FG02--85ER40214.
\newpage
\newcommand{\sAppendix}[1]{%	simplified (starred) form
{\flushright\large\bfseries\appendixname\par
\nohyphens\centering#1\par}%
\vspace{\baselineskip}}
\renewcommand{\theequation}{\mbox{\Alph{section}.\arabic{equation}}}
%\begin{appendix}
\appendix
\section{Appendix}
%\noindent {\Large{\bf Appendix A}}
\setcounter{equation}{0}
                 
In this appendix we present details of our evaluation of the Minkowski space
Fourier 
transforms. Our approach is to define the auxiliary, one dimensional transform,

\begin{equation}
e^{-{\mu^2\over 2}x^2}\bar\Delta (x^2)=\int\limits^\infty_{-\infty} \frac{e^{i\omega
x^2}}{2\pi} G(\omega)dw,
\end{equation}
and its inverse
\begin{equation}
G(\omega)=\int\limits^\infty_{-\infty}e^{-i\omega x^2}e^{-{\mu^2\over 2}x^2}\bar\Delta
(x^2)dx^2.
\end{equation}

\noindent It is understood that $\mu^2>0$ here. 
This procedure can be applied to any function of $x^2$ which has a one
dimensional (in $x^2$) Fourier transform. Substituting Eq. (13) into Eq. (A.2) and
evaluating, we find 
\begin{eqnarray}
G(\omega) & = & \int\limits^\infty_{-\infty}
e^{-x^2(i\omega+{\mu^2\over 2})}\bigg[{1\over 4\pi}\bigg(\delta(x^2)-{m^2\over
2}\theta(x^2) {J_1(m\sqrt{x^2})\over m\sqrt{x^2}}\bigg) \bigg]dx^2\nonumber \\
& = & {1\over 4\pi} - {m^2\over 8\pi}\int\limits^\infty_0
e^{-x^2(i\omega+{\mu^2\over 2})}{J_1(m\sqrt{x^2})\over m\sqrt{x^2}}dx^2\nonumber \\
& = & {1\over 4\pi} e^{-m^2/4(i\omega + {\mu^2\over 2})} \\
\nonumber
\end{eqnarray}
 So the evaluation of the first term in the expression for $\bar B(p^2)$, Eq. (18), now involves the integral
\begin{equation}
\bar B(p^2)= {-m\over 8\pi^2}\int\limits^\infty_{-\infty}d\omega
e^{-m^2/4(i\omega + {\mu^2\over 2})}\int d^4 xe^{i\omega x^2+ip\cdot x},
\end{equation}
where we have exchanged the order of integration in writing this version of Eq.
(18). The $x$--integration is a product of Fresnel integrals that yields
\begin{equation}\int d^4 x e^{i\omega x^2+ip\cdot x} =-i\epsilon
(\omega){\pi^2\over \omega^2}e^{-i{p^2\over 4\omega}}.
\end{equation}
Defining $\nu=4/\omega$, the Fourier transform, $\bar B(p^2)$, can be written
\begin{eqnarray}
\bar B(p^2)/m & = & Im\int\limits^\infty_0 d\nu e^{ip^2\nu -
i\nu m^2 {(1-2i\mu^2\nu)\over (1+4\mu^4\nu^2)}}\nonumber \\
& = & {1\over m^2} \int\limits^\infty_0 dze^{-z^2a/(1+a^2z^2)}\sin\Big[t z
-z/(1+a^2z^2)\Big],\\
\nonumber
\end{eqnarray}

\noindent where $a\equiv 2\mu^2/m^2,$ $t\equiv p^2/m^2$ and $z\equiv \nu m^2 $.
One obtains the Fourier transform of $\tilde B(p^2)$ by simply changing
the sign of $\mu^2$ in (A.6).

To evaluate $\bar B (p^2)$ for $p^2>0$, Eq. (A.6), convergence can be
improved by choosing a contour in the first quadrant in the complex $z$--plane
that is equivalent to the one shown in Eq. (A.6) by Cauchy's theorem. Writing
$z=\rho e^{i\theta}$, $0\leq \theta<\pi/2$, one can recast the representation
of $\bar B(p^2)$ in the form 

\begin{equation}
m\bar B (p^2)=\int\limits^\infty_0 d\rho\Big[\sin
\Big(x\rho\cos\theta+\theta-{\rho\cos\theta\over
D}\Big)e^{-x\rho\sin\theta+\rho{\sin\theta\over D}-{a\rho^2\over D}}\Big],
\end{equation}
\noindent where $x=p^2/m^2$, $a=2\mu^2/m^2$, and
$D=(1-a\rho\sin\theta)^2+a^2\rho^2\cos^2\theta$. The choice $\theta=\pi/4$ is
convenient for evaluating Eq. A.7, which converges significantly faster than
Eq. A.6 for $p^2>0$ and thereby speeds--up the numerical integration. A similar
formula holds for $p^2<0$, where $-{\pi\over 2}<\theta<0$. 
\newpage

%\noindent {\Large{\bf Appendix B}}
\section{Appendix }

\setcounter{equation}{0}

The configuration space solution of the DS equation has the factorizable form
$S(x)=e^{f(x^2)}S_0(x)$ for a {\it class} of models of the vertex, {\it
independent of the gluon propagator's form,} as we outline in this appendix.
The $\delta^4(q)$ propagator is a special example where the Ward identity (8)
defines the DS equation vertex and leads to the factorized solution (11). Let
us see how this result can be generalized to other propagators.

A model for the irreducible
vertex 
that has good symmetry properties\cite{zd} and satisfies the abelian
Ward--Takahashi identity can be written in configuration space as

\begin{equation}
\Gamma_\mu(z;x,y)=F_\mu (y-z, x-z)S^{-1}(x,y)\; .
\end{equation} 

\noindent In Eq. (B.1) $S (x,y)$ is the full fermion propagator, and

\begin{equation}
F_\mu(y-z, x-z) = \int \Big(e^{-iq\cdot (y-z)}-e^{-iq\cdot(x-z)}\Big)
\frac{(x-y)_\mu}{q\cdot(x-y)} \frac{d^4q}{(2\pi)^4} + F^T_\mu\; ,
\end{equation}

\noindent where $\partial F^T_\mu/\partial z_\mu=0$. One can represent the
Fourier transform of $\Gamma_\mu$ in terms of the fermion propagator as
\begin{equation}
\Gamma_\mu(p+k,p)=\frac{\partial}{\partial p^\mu}\int^1_0 S^{-1}(p+\alpha
k)d\alpha + \Gamma_\mu^T(p+k,p).
\end{equation} The first term on the right hand side of Eq. (B.3) is constructed
to satisfy 

\begin{equation}
k^\mu\Gamma_\mu(p+k,p)=S^{-1}(p+k)-S^{-1}(p)\; ,
\end{equation}
where $k^\mu\Gamma_\mu^T(p+k,p)=0$.

While the vertex $\Gamma_\mu$ is convenient for discussion of symmetry
properties of a theory, the study of the fermion DS equation is
generally more convenient when the ``dressed" vertex $\Lambda_\mu$ is used;
\begin{eqnarray}
\Lambda_\mu(p+k,p)\equiv S (p+k)\Gamma_\mu(p+k,p)S(p)\; .
\end{eqnarray}
In terms of $\Lambda_\mu(p+k,k)$, our model looks like
\begin{equation}
\Lambda_\mu(p+k,p)=\frac{\partial}{\partial p^\mu}\int^1_0 S(p+\alpha
k)d\alpha + \Lambda_\mu^T(p+k,p),
\end{equation}
with
\begin{equation}
k^\mu\Lambda_\mu(p+k,p)=S(p)-S(p+k)
\end{equation}
and $k^\mu\Lambda_\mu^T=0$. Since $\Lambda^T_\mu$ vanishes at $k=0$, 
we make the approximation that the longitudinal
term dominates in the infrared region of interest here, and only the first
term in Eq. (B.6) is kept in what follows.\footnote{For the model of Eq. (4),
$\Lambda_\mu^T$ does not contribute anyway, so the first term of Eq. B.6 is the
complete contribution.}

The general form of the DS equation is then, in momentum space,
\begin{equation}
1=(\slash p
-m)S(p)-i\int\frac{d^4k}{(2\pi)^4}\gamma_\mu D^{\mu\nu}(k)\Lambda_\nu(p+k,p),
\end{equation}
where $D^{\mu\nu}(k)$ is the gauge boson propagator.

With $\Lambda_\mu(p+k,p)$ modeled by the first term in Eq.~(B.6), the Fourier
transform of Eq. (B.8) is a pure differential equation, which we write as
\begin{equation}
\delta^4(x)=(i\slash\partial-m)S(x)+\Big[\slash\partial f(x^2)\Big] S(x).
\end{equation}
In Eq. (B.9) we have
\begin{eqnarray}
\frac{\partial}{\partial x^\nu}f(x^2) & = & x_\nu\int^1_0 \Big[d_1
(\alpha^2x^2)+\alpha^2x^2d_2(a^2x^2)\Big] \\                          
& = & 2 x_\nu \dot{f}(x^2),\nonumber \\
\nonumber
\end{eqnarray}
with
\begin{equation}
\dot{f}(x^2)\equiv \frac{d}{dx^2}f(x^2)=\frac{1}{2}\int^1_0
d\alpha\Big[d_1(\alpha^2x^2)+\alpha^2x^2d_2(\alpha^2x^2)\Big].
\end{equation}

In writing Eqs. (B.8)-(10), we have expressed the covariant-gauge, gauge-boson
propagator in the general form
\begin{equation}
D_{\mu\nu}(x)=d_1(x^2)g_{\mu\nu}+d_2(x^2)x_\mu x_\nu.
\end{equation}
The general solution to the DS equation (B.9) has the remarkably simple
factorized form
\begin{equation}
S(x)=e^{if(x^2)}S_0(x).
\end{equation}
The Green function $S_0(x)$ satisfies the free equation
\begin{equation}
(i\slash\partial-m)S_0(x)=\delta^4(x),
\end{equation}
where it is to be understood that arbitrary solutions, $S_H$, to the homogeneous
equation $(i\slash\partial -m)S_H(x)=0$ can always be added to solutions of
Eq. (B.14). Our normalization of $f(x^2)$ is chosen to be $f(0)=1$, so the
solution to Eq. (B.10) can be displayed as
\begin{equation}
f(x^2)=\frac{1}{2}\int\limits^{x^2}_0dx{'^2}\int\limits^1_0 d\alpha
\Big[d_1(\alpha^2 x^{'2})+\alpha^2x{'^2}d_2(\alpha ^2x^{'2})\Big],
\end{equation} in those circumstances where the integral converges.
%\end{appendix}

\newpage
%\def\ref{\par\noindent\hangindent\parindent\hangafter1}

%\centerline{\Large{\bf Figure Captions}}
%\bigskip
%\ref Fig. 1. 
%\medskip
%\ref 
%Fig. 2 
%\medskip
%\ref Fig. 3 

\renewcommand{\baselinestretch}{1.0}
\clearpage
%
%\begin{figure}[p] %default is [tbp] (options=t,p,b,h.  See LaTeX manual p.176)
%  \centering{\ \epsfig{figure=fig1.eps,height=5.7in, angle=-90} }

\begin{figure}[p] %default is [tbp] (options=t,p,b,h.  See LaTeX manual p.176)
 \epsfig{figure=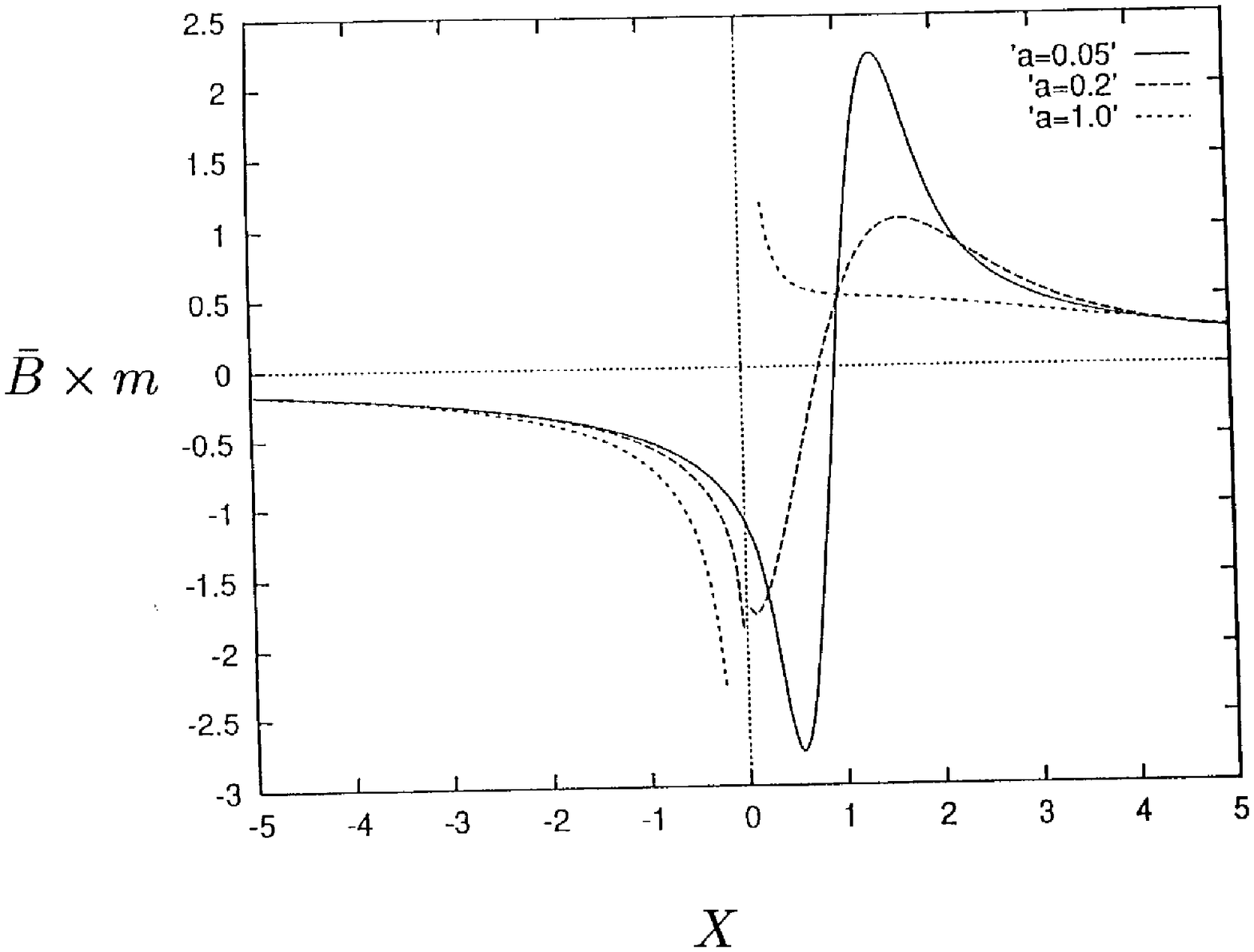,height=5.7in, angle=-90} %}%\parbox{120mm}
\caption{ %PUT CAPTION for figure 1 ON THE NEXT LINE!!!!!!!!!!!
Plot of $m\bar B(p^2)$ vs. $x=p^2/m^2$ for the case $C=0$, Eq.
(21). 
Curves for the values $a=2\mu^2/m^2=0.05$ (solid line), $a=0.2$ (long dashed
line) and $a=1.0$ (short dashed line) are shown. The onset of the pole is
visible for the $a=0.2$ case, 
and it is the dominant feature in the $a=1.0$ case, where the ``resonant"
behavior at $x=1(p^2=m^2)$ is gone. The region around $x=0$ is excluded from
all three plots so that all three cases fit on the figure (The jump across
$x=0$ does not show up on this scale for the $a=0.05$ case). 
\label{fig1} }
\end{figure}
\clearpage
\begin{figure}[p] %default is [tbp] (options=t,p,b,h.  See LaTeX manual p.176)
  \centering{\ \epsfig{figure=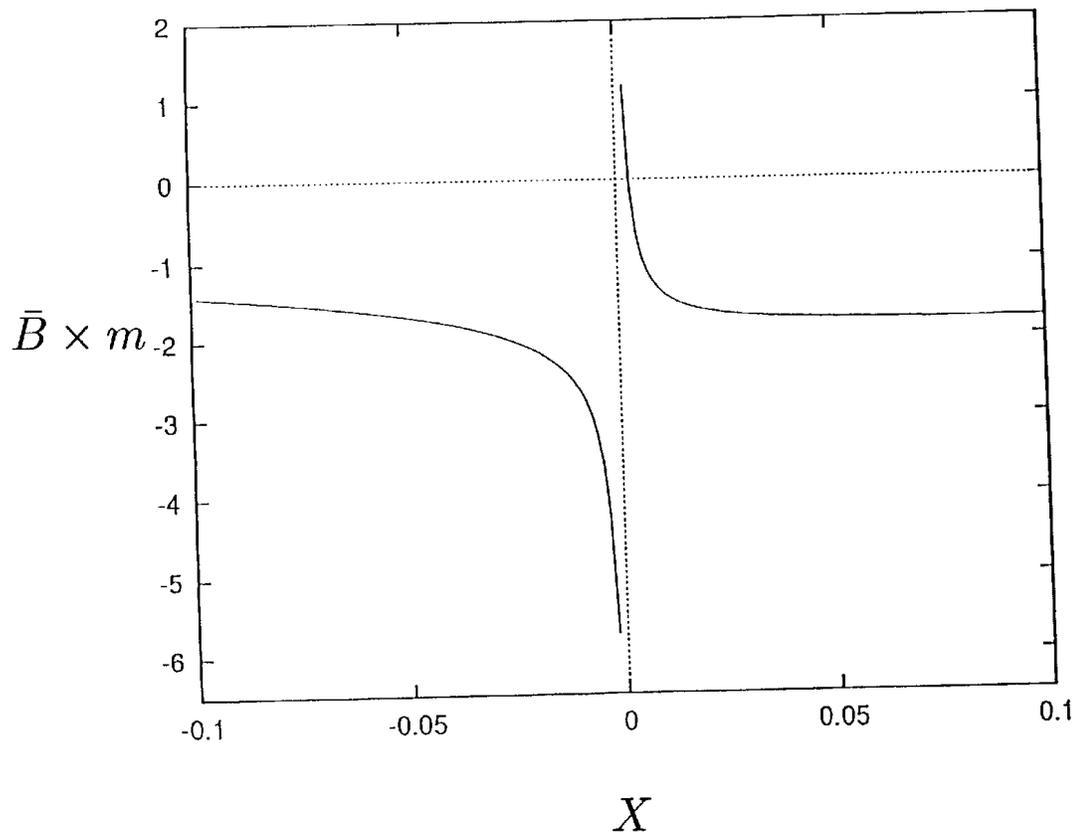,height=5.7in, angle=-90} }
\parbox{120mm}{\caption{ %PUT CAPTION for figure 2 ON THE NEXT LINE!!!!!!!!!!!
A blow--up of the $p^2=0$ region for the $a=0.2$ plot to display
the onset of the singularity. 
\label{fig2} }}
\end{figure}
\clearpage
\begin{figure}[p] %default is [tbp] (options=t,p,b,h.  See LaTeX manual p.176)
  \centering{\ \epsfig{figure=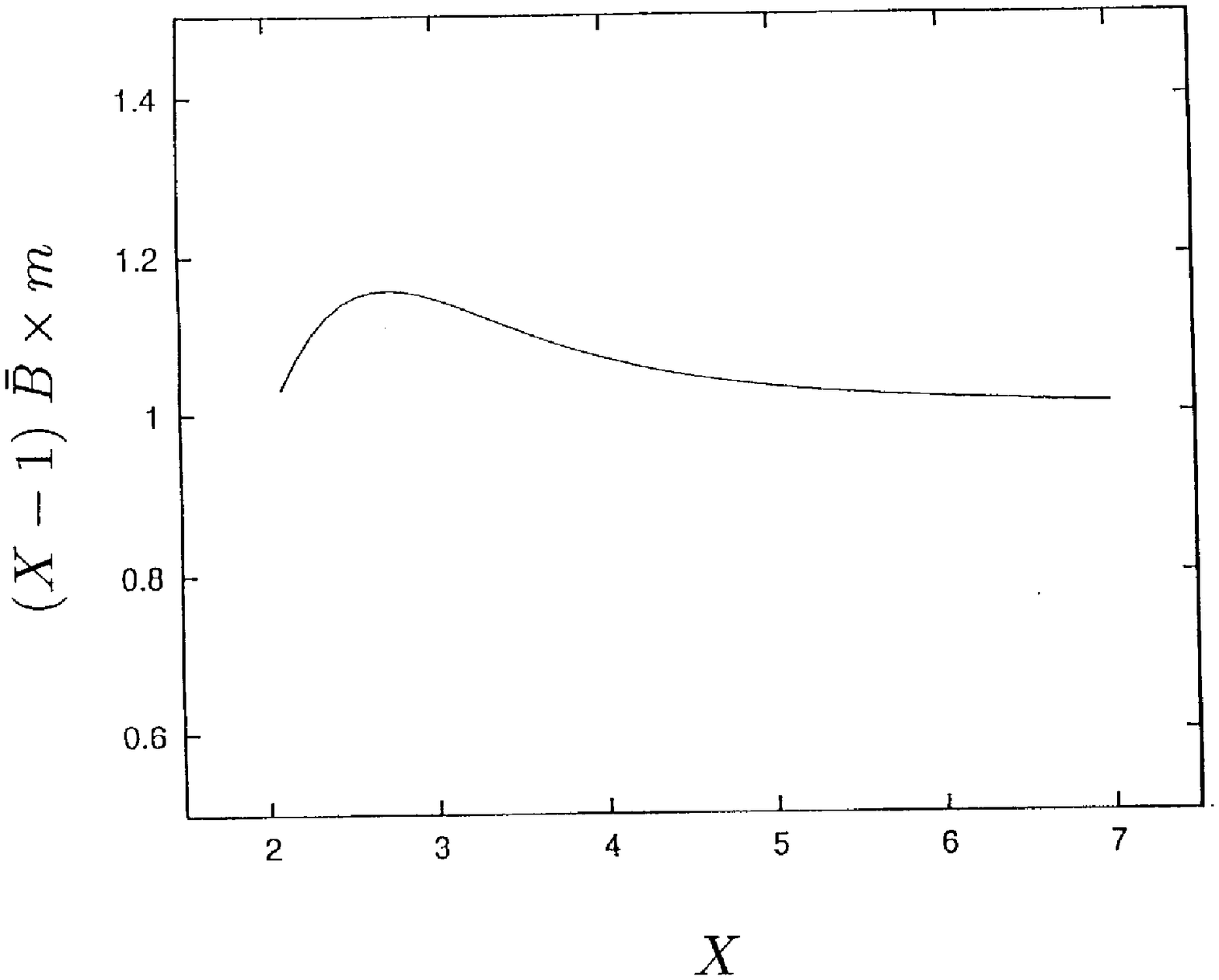,height=5.7in, angle=-90} }
\parbox{120mm}{\caption{ %PUT CAPTION for  fiure3 ON THE NEXT LINE!!!!!!!!!!!
Plot of $(x-1)\bar B(p^2)m$ for the case $a=0.2$. The rapid
approach to 1 as $x=p^2/m^2$ grows, shows the $1/(x-1)$ asymptotic behavior. As
seen in Fig. 1, this behavior is shared by all of the different $a$--value
solutions. The same asymptotic behavior is obeyed by $\tilde B(p^2)$ (see
text).        
\label{fig3} }}
\end{figure}
\newpage


\begin{thebibliography}{99}
\bibitem{prev} Further necessary conditions to satisfy the asympotic condition are
spelled out by Schweber [11], where discussion and references can be found.
\end{thebibliography}
\end{document}